# The Trialkylsulfonium Cation Holds Promise to Capture Carbon Dioxide: In-Silico Evidence Toward a Novel Carbon Dioxide Scavenger

Vitaly V. Chaban[1]

Yerevan State University, Yerevan, 0025, Armenia.

Abstract. The concentration of carbon dioxide ($CO_2$) in the Earth's atmosphere is linked to the acute problem of global warming. For the first time, we herein introduce trialkylsulfonium aprotic ionic liquids (ILs) as a group of seemingly highly capacitive $CO_2$ scavengers. We advocate the viability of the new sorbents by the reaction profiles recorded by means of hybrid density functional theory. All stages constituting $CO_2$ chemisorption, such as the ethyldimethylsulfonium $[S_{211}]$ cation deprotonation, $[S_{211}]$-ylide carboxylation at the α-methylene group, and aprotic anion carboxamidation have been explored. The $[S_{211}]$-ylide formation reaction is thermochemically forbidden but its cost is affordable. The energetic cost ranges from a tiny number of +14 kJ/mol for $[S_{211}]$ indazolide to a mediocre value of +50 kJ/mol for $[S_{211}]$ 1,2,4-triazolide. The barriers corresponding to the sulfonium-based cation deprotonation range from +39 kJ/mol for benzimidazolide to +60 kJ/mol for 1,2,4-triazolide. The energy loss during the ylide intermediate formation is strongly compensated for by the subsequent sulfonium ylide carboxylation. The energetic gain is weakly dependent on the nature of the heterocyclic aprotic anion being -99 to -110 kJ/mol for different ionic species studied. The AHAs additionally participate in $CO_2$ capture following the route of carboxamidation thanks to their nitrogen sites. The most thermochemically favorable carbamate forms out of $[S_{211}]$ indazolide, -68 kJ/mol. The steric and covalent barriers associated with these reactions are of the order of thermal motion energy, whereas a specific chemical structure of AHA engenders marginal differences. The rationalization of the energy reaction profiles is given in terms of partial atomic charges, geometrical peculiarities, steric barriers, imaginary vibrational frequencies, and related descriptors. The reported computational results favor trials of the sulfonium-based ILs for $CO_2$ chemisorption both thermochemically and kinetically.

---

[1] E-mail: vvchaban@gmail.com.

TOC Image

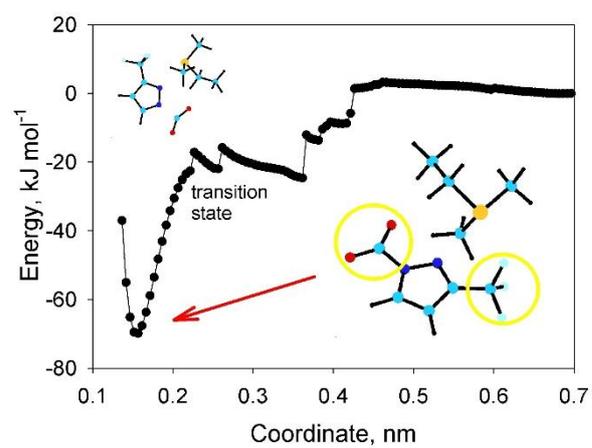

Sulfonium-based ionic liquids reversible produce sulfonium ylide which chemisorbs carbon dioxide.

**Introduction**

Global warming remains a fashionable topic of discussion during the last decades. It readily inspires TV shows, catastrophe movies, and political demarches. Global warming is being linked to the greenhouse gas concentrations in the atmosphere of the Earth by many modern researchers.[1-4] Carbon dioxide ($CO_2$) is seen as a major human civilization-induced greenhouse gas as its huge portions originate from the combustion of fossil fuels. Although the mentioned connection has never been rigorously proven, efforts throughout the globe are underway to develop robust $CO_2$ scavengers.[5-8]

Carbon capture, utilization, and storage (CCUS) designate an extensive area of fundamental and applied research pursued vigorously by naturalists. CCUS unites both engineering efforts to design a new type of green plant and research in chemistry aiming to introduce efficient and ecological $CO_2$ sorbents.[1] The emitted gas originating from a biomass power plant or a chemical plant must be bound before it enters the atmosphere and transported to the location of permanent storage. Carbon sequestration is an approach to storing $CO_2$ for centuries and, possibly, millennia. Carbon dioxide is seen as a chemical with a low intrinsic value as it is highly kinetically stable. CCUS represents a relatively new concept whose purpose is not only to absorb $CO_2$ but also to use a certain amount of it to produce expensive chemicals. In such a way, the expenses associated with $CO_2$ capture can be offset.[9]

Numerous interesting materials and systems have been proposed to bind $CO_2$.[10-12] However, the development of an ideal $CO_2$ scavenger is a challenging endeavor. First, the $CO_2$ molecule represents a global energy minimum for its constituents, i.e., one carbon atom and two oxygen atoms. $CO_2$ is an ultimate product of combustion and cannot be further decomposed into more thermodynamically stable chemical structures. Second, $CO_2$ contains its chemical elements in their most stable degrees of oxidation. Thus, a spontaneous transformation of $CO_2$ into something else should not be expected under even extreme conditions. Third, $CO_2$ features a low normal boiling point which is the result of the inherently weak intermolecular interactions in this substance. Hence, relatively large adsorbent-adsorbate attraction forces are required to keep $CO_2$ captured at finite temperatures.

Room-temperature ionic liquids (RTILs) represent a versatile group of chemical compounds that are composed solely of ions. The ions can be both organic and inorganic. Due to their low volatilities and generally high chemical stabilities, some RTILs are good candidates to be solvents, reaction media, and adsorbents for many small molecules and macromolecules.[8,13-17] Sulfonium-based RTILs are a recently emerged group that has useful physical and chemical properties to find applications in biochemistry, organic synthesis, and electrochemical devices as stable and high-performance electrolytes.[18-22] Sulfonium salts are important precursors to the sulfonium-based ylides, whereas the latter are necessary to engender carbon-carbon bonds. The sulfonium moiety can be used to graft a positive charge to the needed site within a bulky molecule. It is an essential detail in applications that the chosen RTIL had a relatively low shear viscosity to simplify the associated manipulations and eliminate technical hurdles. Shear viscosity can be, to a large extent, modulated by choosing suitable constituents.

Computer simulations represent a powerful and frequently indispensable tool to understand molecular and ion-molecular interactions in many-component systems.[6,17,23] While ab initio electronic-structure methods shed light on the electronic-level phenomena, large-scale molecular dynamics simulations employing empirical potentials are used to observe phase behavior, nanoscale ordering, packing, and collective transport properties.

The concerted analysis of the in-silico generated data combined with conventional chemical wisdom strategically guides the experiment and interprets it.

In this work, we introduce, for the first time, a new group of $CO_2$ sorbents that are based on a sulfonium-based cation and aprotic heterocyclic anions (AHAs). 1,2,4-triazolide [TRIAZ], indazolide [INDA], 3-trifluoromethyl-pyrazolide [3FPYRA], benzimidazolide [BENZIM] were combined with the ethyldimethylsulfonium [$S_{211}$] cation to obtain four prospective ionic $CO_2$ scavengers. Apart from its natural fundamental appeal, the new group of $CO_2$ adsorbents, [$S_{211}$][TRIAZ], [$S_{211}$][INDA], [$S_{211}$][3FPYRA], and [$S_{211}$][BENZIM], can have practical implications under certain conditions.

An acyclic anion, bis(trifluoromethanesulfonyl)imide, [TFSI], was additionally used to quantitatively compare the performance of aromatic and aliphatic, [$S_{211}$][TFSI], structures in deprotonating the α-carbon atom of the sulfonium cation. Hereby we opted to use the sulfonium-based cation with the short alkyl chains to concentrate on the impact of the anion and neglect the steric hindrances. The chemical identity of AHA is deemed important for thermochemistry (total energy gain or loss) and kinetics (the height of the activation barrier) of the ylide formation stage. The ability of the gas molecules to routinely approach the coveted reaction centers of the RTIL to undergo chemical transformations and its dependence on the system's shear viscosity constitutes a cornerstone research topic. The variation of alkyl chain lengths and mutual symmetry is used to engineer solvents and adsorbents with task-specific sets of physicochemical properties.

**Methods, procedures, software, and computed properties**

The series of electronic-structure single-point calculations of the supplied Z-matrices were conducted using the M11 exchange-correlation functional of the hybrid density functional theory.[24] The empirical correction for the dispersion attraction was applied on the fly as an integral part of the M11 Hamiltonian by following each calculation of the converged electronic wave function.[25] The atom-centered split-valence double-zeta polarized basis set 6-31G(d) was employed to construct the molecular wave functions of the simulated systems. The convergence criterion for the molecular wave function was set to $10^{-7}$ Hartree. Where necessary, the system's geometry was optimized using the rational function optimization algorithm until the following convergence criteria were satisfied. The maximum force existing in the simulated system must be brought below 40 kJ mol$^{-1}$ nm$^{-1}$. In turn, the maximum displacement in the system must be brought below $0.5 \times 10^{-3}$ nm. The partial electrostatic charges were computed after the geometry and wave function respective convergence using the Merz-Kollman method to fit the ab initio-derived electrostatic potentials to a set of atomic nuclei.[26]

The study of $CO_2$ chemisorption reaction energy profiles was carried out separately for deprotonation (sulfonium ylide formation), cation carboxylation, and carboxamidation at the chosen anion nitrogen atom. The same site for anion's protonation and carboxamidation was used being the most chemically active, i.e., electron-rich, nitrogen atom of the anion. The independent reaction coordinates were defined as follows: hydrogen (cation)-nitrogen (anion) for the proton transfer stage, carbon (cation)-carbon ($CO_2$) for the carboxylation stage, and carbon($CO_2$)-nitrogen (anion) for the carboxamidation stage. The non-covalent interatomic distances corresponding to the reaction coordinates were scanned continuously to explicitly simulate the chemical phenomena of interest. In general, the reaction coordinate choice can be completely arbitrary as long as it permits the necessary event to occur. The starting molecular configurations for each reaction coordinate scan were obtained by local geometry optimization of the entire system with no restraints imposed. Consequently, the relative potential energy in the first scan point equals zero.

The activation energy is defined as a difference between the highest absolute HDFT-based potential energy in the saddle point and the closest local minimum. Alternatively, an interested reader may use the provided data to readily recompute the barriers versus a physisorption state (the energy at the largest reaction coordinate). The difference is in no case significant. The thermochemical effect of a reaction (ylide formation, $CO_2$ chemisorption at the cation, $CO_2$ chemisorption at the anion) was defined as the difference between the absolute HDFT-based potential energy in the product-related minimum point and the absolute potential energy in point 1. The starting reaction coordinate represents one of a few possible physisorbed states of $CO_2$ in the simulated RTIL. It is obtained via a local geometry optimization procedure applied to randomly placed components.

During the scan of the reaction coordinates, the chosen interatomic separation was stepwise decreased by 0.005 nm corresponding to the gradual approach of the reactants to one another. At every step, the system was allowed to accommodate the change by adjusting its entire Z-matrix excluding the reaction coordinate. Since the system adapts its geometry smoothly, we do not get any unphysical energies and ion-molecular configurations. The potential energy of the system after such a partial geometry optimization was recorded along the coordinate. This lucid algorithm allows us to identify all extrema on the potential energy surfaces along the chosen directions of movement and link them to specific chemical events. The reaction coordinate scan continues several points beyond the formation of products. I.e., the interatomic distance decreases down to the respective covalent bond length. Due to the complicated potential energy surface of the simulated many-atom systems, the positions of maxima and minima may appear somewhat shifted compared to the simplified models.

The identities of the found stationary points, a minimum or transition state, were verified by computing the profiles of vibrational frequencies according to the conventional harmonic rigid rotor approximation. All real-number frequencies were an indication that a point of minimum was located. In turn, the presence of a single imaginary frequency among the real-number frequencies meant that a first-order saddle point was obtained.

The study of the reaction profiles represents a computationally intensive time-consuming endeavor that anticipates thousands of consequent self-consistent field calculations. The simulated systems evolve gradually following the forcibly moved reaction coordinates and thus mimic steered molecular dynamics simulations. In the present work, the calculations were executed in parallel by involving 64 threads per procedure.

GAMESS.2020 was used to conduct electronic-structure calculations.[27] Avogadro 1.2.0 was used to prepare initial Z-matrices.[28] Gabedit 2.5.1 was used to inspect the output structures and summarize necessary information.[29] The functions and procedures defined in the scientific libraries SciPY.org[30] and ASE[31] were used to manipulate the streams of chemical data.

**Design of simulations and interpretation of reaction profiles**

Figure 1 depicts the structural chemical formulas of ions that participated in the simulated $CO_2$ chemisorption reactions. To construct suitable adsorbents for $CO_2$, four nitrogen-containing AHAs (Figure 1A-D) were coupled with the ethyldimethylsulfonium cation (Figure 1F). As a result of the $CO_2$ chemisorption reaction, α-carboxyl ethyldimethylsulfonium (Figure 1G) forms. The obtained compound represents a fundamentally interesting zwitterion, in which the positive and negative charges are minimally spatially separated. As a result of the small interatomic distance, both charges are essentially delocalized (vide infra). The intramolecular electrostatic attraction between the mentioned interaction centers is an unusual structural pattern that additionally stabilizes such a product of $CO_2$ chemisorption. The acyclic nitrogen-containing anion, bis(trifluoromethanesulfonyl)imide (Figure 1E), was added to the investigated set to separate the impact of the anion's aromatic structure. Therefore, the role of the high-energy π-electronic density of AHAs in the deprotonation of the sulfonium-based cation, i.e., the formation of the sulfonium-based ylide, can be unraveled.

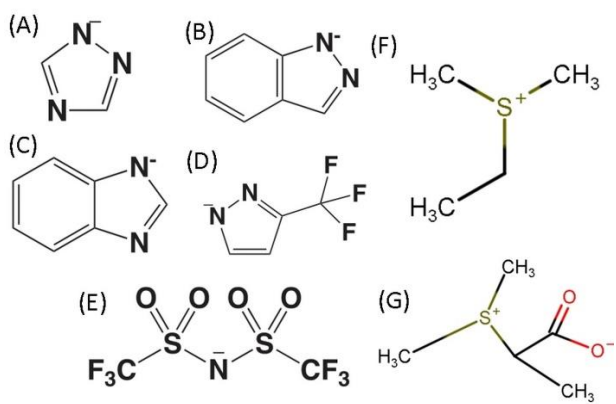

Figure 1. The simulated models of the sulfonium-based cation and several anions. (A) 1,2,4-triazolide; (B) indazolide; (C) benzimidazolide; (D) 3-trifluoromethyl-pyrazolide; (E) bis(trifluoromethanesulfonyl)imide; (E) ethyldimethylsulfonium; (G) carboxylated ethyldimethylsulfonium.

Every chemical transformation in a condensed phase deals with two kinds of activation barriers. First, the reagents must appear in the vicinity of one another. This means that non-covalent distances of roughly 130-150% of the prospective covalent bond lengths must be attained. In physical systems, the trajectories of the interacting particles intersect randomly and sporadically during their thermal motion under a given temperature and pressure.[32]

Potential barriers routinely exist in the condensed-phase systems preventing one moiety from colliding with another moiety. For instance, in [S$_{211}$][AHA]s, an anion tightly coordinates the positively charged region of the cation and, therefore, blocks its direct collision with $CO_2$ molecules. The alkyl chains of the trialkylsulfonium cation also represent an obstacle faced by the gas molecules. They ultimately deteriorate the rate of the chemisorption reaction. The reacting particles or regions of bulky particles must overcome the mentioned barriers thanks to their kinetic energy at a given temperature prior to considering the possibility of a chemical transformation. We herein refer to such energetic costs as steric barriers. The steric barriers are, on average, relatively small being on the order of a dozen of kJ mol$^{-1}$. Their specific heights depend on the intensity of intermolecular interactions in a given region of a system but may differ drastically from case to case.

Another type of activation barrier is the covalent one. These barriers are associated with the covalent electron-electron electrostatic repulsion and attraction. The heights of the covalent activation barriers may be significantly larger as compared to that of the above-discussed steric barriers. Indeed, electron-nucleus and electron-electron interaction energies are higher than the dispersion and long-range Coulombic energies. In rare cases, the chemical reaction proceeds with no or tiny covalent barriers though. For example, such processes can be observed when a reaction intermediate represents a metastable chemical entity. Furthermore, the actual height of the covalent barrier depends on the molecular environment in a given region of the configurational space. The $CO_2$ chemisorption reactions involving sulfonium-based RTILs are modulated by the chemical identity of the anion although the anion does not formally participate in the carboxylation of the sulfonium-based cation. Understanding the precise mechanism of such reactions is necessary to reveal their bottlenecks and step-by-step thermochemistry. This knowledge allows one to adjust the laboratory setups whenever possible to attain higher product yields and more technologically and economically feasible chemisorption conditions.

**Results and discussion**

**Trialkylsulfonium-based ylide formation from trialkylsulfonium cation**

Figure 2 investigates the deprotonation of the [$S_{211}$] cation and the subsequent protonation of the employed AHAs. As a result of the cationic deprotonation, the sulfonium-based ylide emerges. The latter is a relatively stable chemical compound that contains a double bond between the sulfur atom and the α-carbon atom. The sulfonium-based cations and the corresponding sulfonium-based ylides possess quite different structures in terms of geometry (covalent bonds, covalent angles, covalent dihedrals, vide infra) and electron density distribution. Thanks to its double sulfur-carbon bond, the sulfonium-based ylides are much more chemically reactive than the corresponding sulfonium-based cations. As we recently showed numerically, namely the emergence of ylide makes it possible to fulfill the thermodynamically and kinetically allowed carboxylation of the quaternary cations.[21,23]

In the case of all investigated AHAs, the proton transfer stage leading to the formation of the sulfonium-based ylide is associated with the clearly defined covalent potential barriers. The heights of the barriers depend on the chemical nature of the anion and range between 40 and 60 kJ mol$^{-1}$ (Figure 2). In turn, the steric potential barriers do not apply to the deprotonation reactions because of small interionic equilibrium distances in all systems. Note that the distance separating the most electron-rich interaction center of the anion (nitrogen atom) and the hydrogen atom of the sulfonium α-methylene group does not exceed 0.25 nm in the case of any of the investigated [$S_{211}$][AHA] RTILs. Due to such structural patterns in the condensed state, the cation-anion proton transfer needs to cope with a single obstacle. As Brennecke and coworkers showed in their interesting works on the phosphonium-based cations, moderate heating is required to activate the formation of the ylide and launch the subsequent $CO_2$ chemisorption stages.[33] Our present results indicate that a similar requirement also applies to the sulfonium-based cations.

The proton transfers are thermochemically and kinetically hindered according to the performed HDFT-powered calculations. The highest cost is seen in [$S_{211}$][TRIAZ], 50 kJ/mol, whereas the lowest cost, 13 kJ/mol, is seen in [$S_{211}$][INDA]. Indeed, the [$S_{211}$]-based cation is a more stable structure than the corresponding ylide. Note that the internal energy reaction effects are given relative to the non-reacted states of each RTIL. Therefore, the provided energies can be directly used in the calculations of the total energetic effects of the investigated $CO_2$ chemisorption reactions.

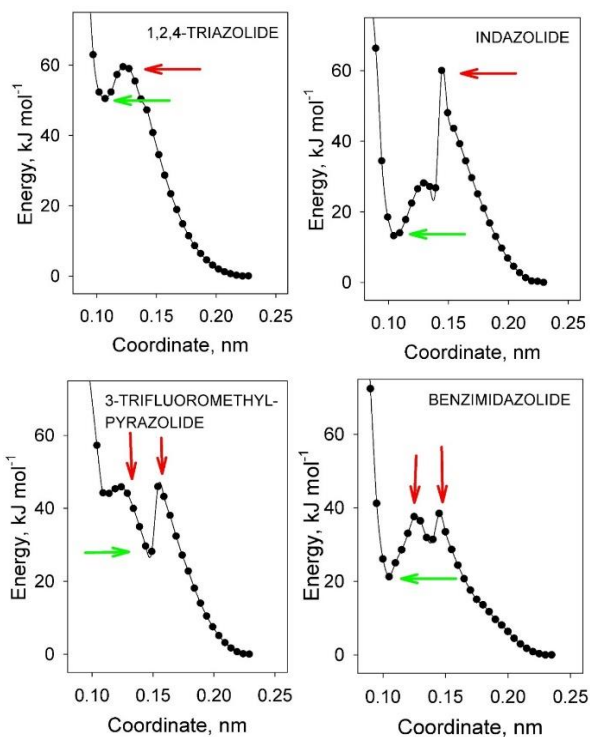

Figure 2. The evolution of the reaction coordinates (nitrogen-hydrogen distances) corresponds to the proton transfer in the chosen [$S_{211}$]-based RTILs, see legends. The arrows depict the most representative maxima and minima.

The heights of the activation barriers correlate with thermochemistry in three out of four RTILs. [$S_{211}$][INDA] represents an exception wherein the barrier of 61 kJ/mol coexists with a comparatively favorable thermochemistry of just +13 kJ/mol. Such a drastic difference implies that [$S_{211}$][INDA] adjusts its configuration after the saddle point is passed by the transferred proton. The changes are induced by the transformation of the COO moieties from a linear electrostatically neutral structure to a negatively charged carboxylate group.

The maxima in Figure 2 correspond to the transition states as it was verified by the analyses of vibrational frequency profiles. A single imaginary frequency per profile was obtained for the evaluated maximum points: i907 cm$^{-1}$ in [$S_{211}$][TRIAZ], i418 cm$^{-1}$ in [$S_{211}$][INDA], i973 cm$^{-1}$ in [$S_{211}$][3FPYRA], and i1278 cm$^{-1}$ in [$S_{211}$][BENZIM]. The magnitudes of the imaginary vibrational frequencies indicate the steepness of the slopes on the potential energy surface in the proximity of the maxima. Such steepnesses appear very different depending on the nature of the AHA. To record an effect of the employed basis set on the imaginary frequencies, we conducted an additional series of calculations for [$S_{211}$][TRIAZ] using a substantially more comprehensive set of basis functions, 6-311++G**, that includes diffuse and polarization functions on even the hydrogen atoms. The resulting imaginary frequency appeared 20 cm$^{-1}$, i.e., two percent, smaller. Therefore, 6-31G(d) provides a competitively accurate description of the simulated systems and phenomena.

The nitrogen-hydrogen distance in the deprotonation transition state detected in [$S_{211}$][TRIAZ] amounts to 0.122 nm (Figure 3) which is 24% smaller than the carbon-hydrogen distance, 0.151 nm. Thus, the deprotonation transition state is closer to the product state. The elongation of the carbon-hydrogen bond that breaks apart at the transition state is 39%. The locations of the bond-breaking transition states at such covalent bond elongations represent a conventional observation.[32,34] The carbon-hydrogen-nitrogen angle in the transition state equals 163 degrees and, therefore, somewhat deviates from the straight angle.

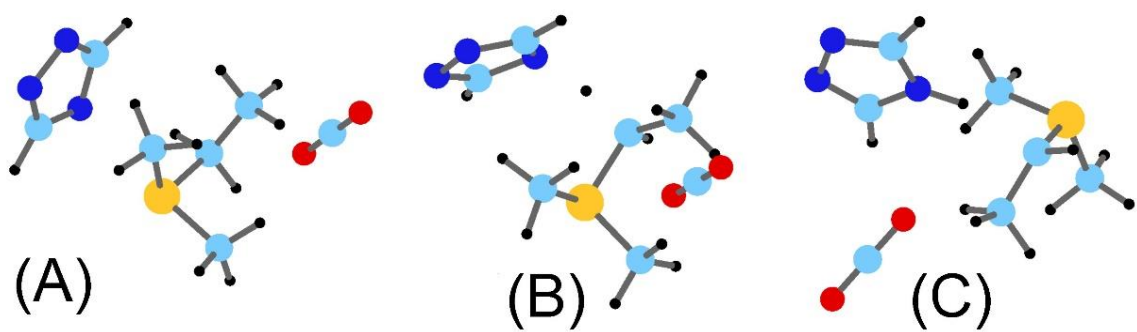

Figure 3. The representative configurations correspond to $CO_2$ chemisorption by [$S_{211}$][TRIAZ]. (A) [$S_{211}$] and [TRIAZ]. (B) Transition state for the hydrogen atom transferring from [$S_{211}$] to [TRIAZ]. (C) [$S_{211}$]-based ylide, and [H][TRIAZ]. The carbon atoms are cyan, the hydrogen atoms are black, the nitrogen atoms are blue, the oxygen atoms are red, and the sulfur atom is yellow.

Other [S$_{211}$][AHA]s exhibit similar geometric features (Figure 4). For instance, the carbon-hydrogen distances characterizing the ylide formation transition states are 0.144 nm in [S$_{211}$][INDA], 0.149 nm in [S$_{211}$][3FPYRA], and 0.147 nm in [S$_{211}$][BENZIM]. The nitrogen-hydrogen distances, in turn, amount to 0.125 nm in [S$_{211}$][INDA], 0.124 nm in [S$_{211}$][3FPYRA], and 0.125 nm in [S$_{211}$][BENZIM]. The carbon-hydrogen-nitrogen angles are 158 degrees in [S$_{211}$][INDA], 160 degrees in [S$_{211}$][3FPYRA], and 163 degrees in [S$_{211}$][BENZIM].

Noteworthy, two maxima points correspond to deprotonation in all three RTILs and reflect the symmetry of the electron-rich sites within the AHA. In the [3FPYRA] and [BENZIM] anions, the symmetry of the nitrogen atoms manifests out in the reaction profile particularly strongly. The carbon-hydrogen and nitrogen-hydrogen distances swap in the additional transition states, e.g., r(C-H) gets to 0.123 nm and r(N-H) gets to 0.147 nm in [S$_{211}$][BENZIM]. Note that we simulated the reaction for the most nucleophilic nitrogen atom of AHA, which was determined according to the partial charge distributions in the ionic state of the sorbent.

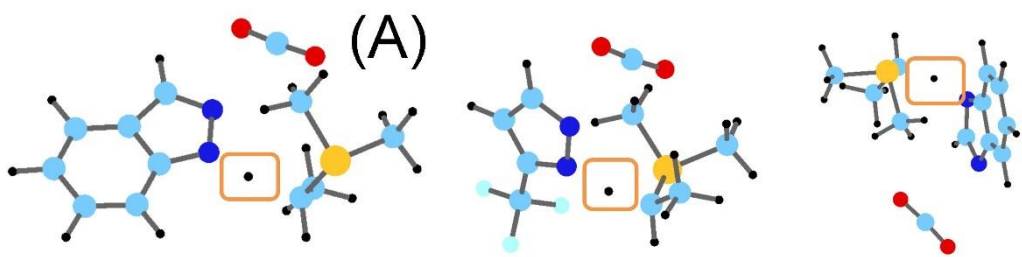

Figure 4. The transition state molecular geometries corresponding to the cation-anion proton transfer processes in (A) [S$_{211}$][INDA]; (B) [S$_{211}$][3FPYRA]; (C) [S$_{211}$][BENZIM]. The orange rectangles underline the immediate positions of the migrating hydrogen atoms in the maxima points on each potential energy surface. The carbon atoms are cyan, the hydrogen atoms are black, the nitrogen atoms are blue, the oxygen atoms are red, the fluorine atoms are light blue, and the sulfur atom is yellow.

The anion's geometry appears insensitive to proton acceptance. The two carbon-nitrogen covalent bonds of the [TRIAZ] anion, wherein nitrogen is a proton acceptor, are 0.135 nm long in the ionic state and 0.136 nm long in the neutral (protonated) state. The carbon-nitrogen-carbon covalent angle in this anion equals 100 degrees in the ionic state, 102 degrees in the transition state, and 104 degrees in the protonated state. Within the ethyldimethylsulfonium cation, the alterations of the covalent bond lengths are more characteristic. The carbon-hydrogen bond length in the methine group of the sulfonium-based ylide shrinks down to 0.101 nm, whereas it equals 0.110 nm in the sulfonium-based cation. The sulfur-carbon bond length shrinks down to 0.176 nm versus 0.184 nm in [$S_{211}$]. However, the sulfur-carbon-carbon covalent angle changes insignificantly, from 115 to 114 degrees.

The carbon dioxide molecule does not play an essential role in the cation-anion proton exchange. One may corroborate such a conclusion using the provided visualizations (Figures 3-4). For instance, the closest-approach $CO_2$-cation distance during the entire proton exchange reaction is 0.264 nm. The corresponding distance was measured between the oxygen atom of $CO_2$ and the hydrogen atom of the methyl group of one of the three accessible alkyl chains. We underline that even though $CO_2$ approaches the terminating atom of the alkyl chain, it does not diffuse toward the emerging methine moiety.

The $CO_2$-anion distance is even larger. The distance found between the hydrogen atom attached to one of the carbon atoms of [AHA] and the oxygen atom of $CO_2$ in the transition state geometry amounts to 0.444 nm. The closest-approach distance of $CO_2$ to the migrating proton is 0.492 nm. There is no direct contact between $CO_2$ and the reactants participating in the proton transfer.

The $CO_2$ molecule remains insensitive to the proton exchange. It does not undergo any structural deformations and the partial charges change marginally. In the ionic state of the sorbent, q(C)=+0.70e, and q($O_{1,2}$)=-0.35e, whereas q(C)=+0.72e, and q($O_{1,2}$)=-0.36e in the ylide state of the sorbent.

The stepwise electron density redistribution during the chemical reaction can be analyzed in terms of atomic nucleophilicities. The nucleophilicity can be quantified via positive atomic changes. The latter, in turn, can be computed from the ab initio-derived molecular electrostatic potential. We hereby provide several most descriptive charges corresponding to the ionic state, transition state, and ylide state. The most significant alterations in the electron density occur at the nitrogen atoms of the anions, the α-carbon atom of the cation, and, naturally, the traveling proton.

In the ionic state of [$S_{211}$][TRIAZ], the partial charge of the methylene hydrogen atom equals +0.07e, while the partial charge of the most nucleophilic anion's nitrogen atom equals -0.76e. The center of the cation is just slightly electrophilic, +0.15e. Once the [TRIAZ] anion protonates, the amount of electron density on an initially strong nucleophile, the nitrogen atom, decreases drastically, to +0.23e. The charge of the transferred hydrogen atom becomes -0.15e. The charge of the center of ethyldimethylsulfonium ylide, a sulfur atom, becomes +0.05e. The charge of the methine carbon atom becomes -0.13e versus +0.04e in the cationic state. In the transition state, the charges distribute as follows: -0.11e (nitrogen), +0.26e (α-carbon), and -0.18e (migrating hydrogen).

The partial charges fitted from the ab initio-derived electrostatic potential are highly sensitive to the molecular environment. They may differ essentially at each step of the considered chemical reaction. In the meantime, such charges are important descriptors of the electronic density evolution during the reaction and provide specific information regarding nucleophilic and electrophilic interaction sites at every step of the process.

Unlike [TRIAZ] which features three nitrogens, the [INDA], [3FPYRA], and [BENZIM] contain two nitrogen atoms. These nitrogen atoms represent the most nucleophilic reaction sites in the corresponding [S$_{211}$][AHA]s. Therefore, they are involved in all $CO_2$ chemisorption-associated processes. In the ionic states, the nucleophilicities of the nitrogen atoms amount to -0.52e and -0.27e in [INDA], -0.39e and -0.40e in [3DPYRA], -0.75e and -0.79e in [BENZIM]. Therefore, all nitrogen atoms are essentially nucleophilic in [S$_{221}$][AHA]s. Moreover, the physisorption of $CO_2$ does not impact the exemplified electron density distribution. In the simulations, we assume that the proton opts to chemically bind a more nucleophilic nitrogen atom since it appears closer to α-carbon upon the cation-anion coupling in the equilibrium state.

In the transition states, the nucleophilicities of the nitrogen atoms which accept the proton adjust drastically, e.g., -0.05e in [INDA], -0.08e in [3FPYRA], and -0.37e in [BENZIM]. In the meantime, the nucleophilicities of the neighboring nitrogen atoms remain intact. The product state assumes that the cation exists in the form of the sulfonium-based ylide, i.e., a neutral molecule. Whereas, the AHA transforms into a heterocyclic molecule. Consequently, the nitrogen atom in [H][INDA] gets a charge of +0.27e which is very similar to the case of [H][TRIAZ], see the discussion above. Similarly, the proton-bearing nitrogen atom of [H][3FPYRA] attains electrophilicity, +0.63e. Such a relatively large positive charge originates from a joint action of protonation of the presence of the trifluoromethyl moiety nearby. The only case in which the nitrogen atom retains a slight nucleophilicity after protonation is [H][BENZIM], -0.17e. While the characterized nucleophilicities represent important descriptors of the chemical systems and chemical phenomena, we were unable to identify any clear correlations between the energy evolutions in the reaction profiles and atomic nucleophilicities of the original ionic structures. One must conclude that the formation of the [$S_{211}$]-based ylide depends on numerous factors. They cannot be readily predicted before explicit electronic-structure-powered sampling of the reaction pathways.

It may be interesting to note that our analysis reveals routinely small electrostatic charges on the hydrogen atom that leaves the cation and joins the anion. Particularly, the hydrogen's partial charge decreases from +0.07e to -0.15e for the case of [$S_{211}$][TRIAZ] as this atom leaves the cation, passes the transition state, and eventually joins the AHA. Although we formally term the migrating hydrogen as a proton, it exhibits the electrostatic behavior of a somewhat polarized atom. Nonetheless, the deprotonation reaction fosters a substantial electron density redistribution the result of which is neutralization of [$S_{211}$] and [H][AHA] as separate chemical entities.

The 1,2,4-triazolide anion features a sophisticated valence electron density distribution. The protonating hydrogen atom is not a pronounced electron-deficient center in the conjugated electronic system. Instead, a larger fraction of the positive charge compensates for the excessive electron density on the nitrogen atom of AHA. During the protonation, the partial charge on nitrogen increases from -0.76e in [AHA] to +0.23e in [H][AHA]. Therefore, protonation quenches the affinity of one of the triazolide-forming nitrogen atoms to the counterion.

By accepting the proton, the [TRIAZ] anion becomes a heterocyclic molecule. During the hydrogen atom detachment, the α-carbon atom increases its partial charge gradually, i.e. a significant fraction of the hydrogen's valence electron density leaves the cation. Compare the α-carbon's charge of +0.04e at the carbon-hydrogen distance of 0.110 nm to carbon's charge of +0.43e at the carbon-hydrogen distance of 0.134 nm. The further movement of the hydrogen atom to the nitrogen atom of the anion leads to a decrease in the α-carbon atom electrostatic charge, from +0.42e to -0.11e in the sulfonium-based ylide.

The 1,2,4-triazolide, indazolide, 3-trifluoromethylpyrazolide, and benzimidazolide anions transform into 1,2,4-triazole [H][TRIAZ], indazole [H][INDA], 3-trifluoromethylpyrazole [H][3FPYRA], and benzimidazole [H][BENZIM]. Since the thermochemistry of deprotonation is above zero, one would expect the reaction to be reversible with a larger amount of the [$S_{211}$]-based RTIL remaining in its ionic form. At certain conditions, the ethydimethylsulfonium ylide exists in equilibrium with [$S_{211}$]. Once the carboxylation product emerges and the reaction equilibrium shifts rightward, a new portion of the ylide forms out of the cation. As a result, one may expect a significant alteration in the physicochemical properties of the reaction media. It must be reiterated that the [$S_{211}$]-based ylide plus a heterocyclic molecule lie higher on the potential energy surface as compared to the ionic liquid precursor according to the computed reaction profiles. The subsequent $CO_2$ chemisorption may only be a thermochemically, and thus technologically, feasible process if the next reaction stage fully compensates for the hereby recorded energetic losses.

All investigated [$S_{211}$][AHA]s exhibit largely similar behavior when transforming into the ylide and heterocycle. However, the quantitative differences occur due to the various electron distribution patterns in each AHA. A more nucleophilic nitrogen atom generally fosters more favorable energy alterations. For instance, the most negative charge in [BENZIM] corresponds to a relatively small energy penalty and a relatively low covalent reaction barrier. As we highlighted above, it is impossible to rate the RTILs based exclusively on this parameter. The potential energy surfaces in the aromatic systems containing a few non-carbon atoms are very sophisticated. The high-energy valence electrons are distributed over the aromatic rings, which are very sensitive to every conformational change.

The energetic effect and activation barrier depend not only on the three interaction centers directly participating in it (carbon of cation, migrating hydrogen, and nitrogen of anion) but on the charge density distribution over the entire ion pair, particularly, over the AHA. Despite applying comprehensive efforts to find a unique descriptor to link to the reaction energy effect and the associated activation barrier, it has not been possible to determine a single one. In this context, ab initio-based reaction profiles mimicking the entire flow of the chemical reaction look like an invaluable instrument to monitor the accumulated effect and predict the feasibility of the reaction for each set of reactants and hypothetical products.

**Carboxylation of sulfonium-based ylide**

As soon as the $[S_{211}]$-based ylide forms, the $CO_2$ chemisorption reaction starts. The sulfonium-based ylide represents quite a reactive compound thanks to its double sulfur-carbon covalent bond and the excessive electron density localized nearby. The chemisorption by the ethyldimethylsulfonium ylide is a facile compound reaction during which the $CO_2$ molecule carboxylates the α-carbon atom of the $[S_{211}]$ cation. The resulting α-carboxylated ethyldimethylsulfonium represents a zwitterion. Therein, a positively charged center is sulfur, whereas a negatively charged center is the deprotonated carboxyl group. Since the electron-deficient and electron-rich centers are located nearby, they strongly interact electrostatically. In this way, the zwitterionic product of chemisorption gets additionally stabilized.

Figure 5 summarizes the reaction profiles corresponding to the carboxylation of the ethyldimethylsulfonium ylide in various RTILs. We opted to perform an electronic-structure treatment of all components simultaneously to identify the passive and active roles of the reactants. For instance, [H][AHA] presumably does not participate in the carboxylation stage occurring at the α-carbon atom of the $[S_{211}]$-ylide. Nevertheless, the cation-anion interactions represent a paramount factor determining the local structures of RTILs. The inclusion of the AHA in the research of ylide carboxylation is deemed to be an essential advantage of the employed model.

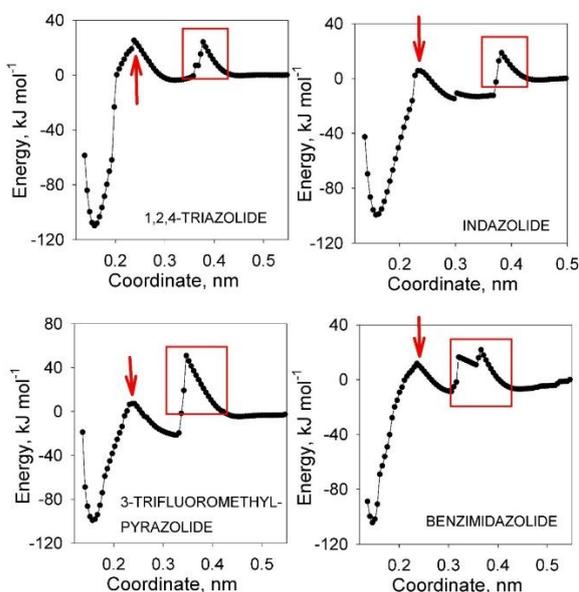

Figure 5. The evolution of the reaction coordinates (carbon of $[S_{211}]$-carbon of $CO_2$ distances) corresponds to the ylide carboxylation in the chosen ethyldimethylsulfonium RTILs, see legends. The red rectangles point to the steric barriers. The red arrows designate covalent barriers. The abrupt changes in energy correspond to the rotations of the reacting particles relative to one another.

The carboxylation of the [$S_{211}$]-ylide may be theorized as a competition between $CO_2$ and [H][AHA] for the location near the most electron-rich interaction site of the ylide which is the α-methine group. The [H][AHA] molecule acts as an obstacle for $CO_2$ approaching the α-methylene moiety. It engenders one or more steric barriers as depicted in Figure 5. In all sorbents, such steric barriers are found at the reaction coordinate values between 0.30 and 0.40 nm. The barriers are linked to the necessity of the $CO_2$ molecule to substitute [H][AHA] in the vicinity of the center of the [$S_{211}$]-ylide (Figure 6). The ylide-[H][AHA] attraction is expected to be somewhat stronger than the ylide-$CO_2$ attraction because the partial charges in heterocycles exceed the partial charges in $CO_2$ as follows from the discussion above. Subsequently, a direct approach of $CO_2$ to the reaction site is challenging. All recorded reaction profiles reflect the above-discussed reaction hurdles in the form of steric barriers.

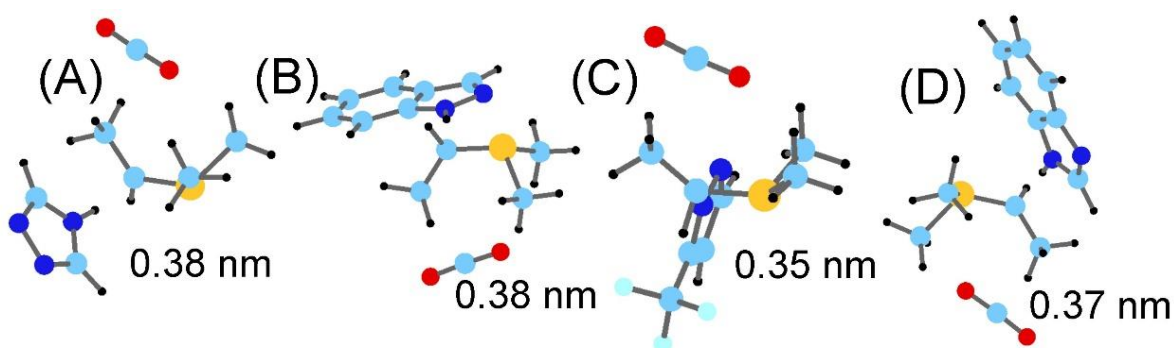

Figure 6. The molecular configurations corresponding to the identified steric barriers in various reacting systems of (A) [$S_{211}$][TRIAZ]; (B) [$S_{211}$][INDA], (C) [$S_{211}$][3FPYRA], and (D) [$S_{211}$][BENZIM]. The steric barriers originate from the competition between $CO_2$ and anions for the location near the α-methine group of the [$S_{211}$]-ylide. The provided distances specify the positions of the detected steric barriers along the sampled reaction coordinate. The carbon atoms are cyan, the hydrogen atoms are black, the nitrogen atoms are blue, the oxygen atoms are red, the fluorine atoms are light blue, and the sulfur atom is yellow.

The revealed steric barriers are several to a few dozen kJ/mol high. In the case of [$S_{211}$][3FPYRA], the steric barrier exceeds 40 kJ/mol. The cause of the barrier is a rotation of $CO_2$ upon a search of the minimum-energy pathway to the α-methine moiety. In the case of [$S_{211}$][BENZIM], there are two adjacent steric barriers of nearly the same height mimicking the symmetry of the heterocyclic structure. The geometry of $CO_2$ does not react substantially to the faced steric barriers. For instance, the covalent oxygen-carbon-oxygen angles of $CO_2$ in the steric transition states equal 175-177 degrees. The covalent C=O bond lengths shrink to 0.117 nm versus 0.118 nm in the gaseous phase of $CO_2$.

The $CO_2$ molecule smoothly grafts to the α-carbon atom of the [$S_{211}$]-ylide thanks to the existence of the above-described S=C double bond (Figure 7). The transition states corresponding to this chemical reaction locate at the following carbon (ylide)-carbon ($CO_2$) distances: 0.238 nm in [$S_{211}$][TRIAZ]; 0.238 nm in [$S_{211}$][INDA], 0.232 nm in [$S_{211}$][3FPYRA], and 0.241 nm in [$S_{211}$][BENZIM]. The imaginary vibrational frequency characterizing the transition state of the sulfonium-based ylide carboxylation reaction in [$S_{211}$][INDA] equals i260 $cm^{-1}$.

The geometry of $CO_2$ in the carboxylation transition state adjusts substantially. The molecule bends so that its valence angle becomes 163 degrees. Compare this to the equilibrium value of 180 degrees in gaseous $CO_2$. In the carboxylated [$S_{211}$]-ylide, the oxygen-carbon-oxygen angle equals 132 degrees. In turn, both carbon-oxygen distances change insignificantly in the transition state, 0.118 nm. In the carboxylated [$S_{211}$]-ylide state, the discussed distances are 0.124 nm, whereas gaseous $CO_2$ exhibits a distance of 0.117 nm. The lengthening of covalent bonds in $CO_2$ manifests the formation of the electron-rich carboxylate moiety.

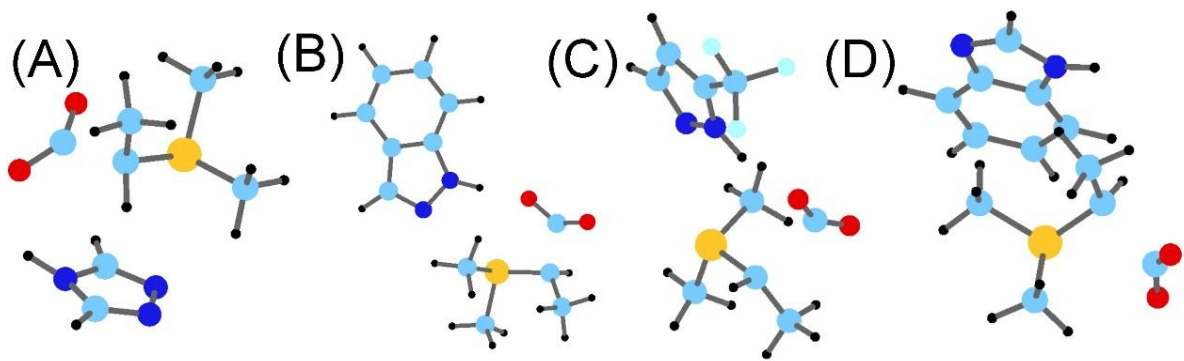

Figure 7. The deformations experienced by the $CO_2$ molecule during the $[S_{211}]$-ylide carboxylation reaction in various investigated RTILs: (A) $[S_{211}][TRIAZ]$; (B) $[S_{211}][INDA]$, (C) $[S_{211}][3FPYRA]$, and (D) $[S_{211}][BENZ]$. The carbon atoms are cyan, the hydrogen atoms are black, the nitrogen atoms are blue, the oxygen atoms are red, the fluorine atoms are light blue, and the sulfur atom is yellow.

Figure 8 depicts the relaxed geometries of the carboxylated ethyldimethylsulfonium zwitterion. In this state, the structure retains electrical neutrality, whereas the heterocycles exist as molecules. We anticipate that neutral particles coexist with ions in real-world systems because the chemisorption reaction is expected to be reversible. The carboxylation of the ylide is a thermochemically favorable process. It brings to the system roughly -100 kJ/mol in each of the four $[S_{211}][AHA]$ RTILs. In the state of ylide, the α-carbon atom coordinates the electron-rich nitrogen atom of [H][AHA]. In turn, the carboxylated sulfonium-based zwitterion coordinates the heterocyclic molecule via the carboxyl moiety. The mentioned interaction is a central structural peculiarity in all combinations of $[S_{211}]×CO_2$ and heterocyclic moieties. The potential energy surface favors such a pattern adjustment due to a stabilizing impact on the products of $CO_2$ chemisorption. The intermolecular oxygen-hydrogen distances amount to 0.207 nm in $[S_{211}][TRIAZ]$, 0.179 nm in $[S_{211}][INDA]$, 0.170 nm in $[S_{211}][3FPYRA]$, and 0.180 nm in $[S_{211}][BENZIM]$. The O…H-N angles amount to 124 degrees in $[S_{211}][TRIAZ]$, 172 degrees in $[S_{211}][INDA]$, 152 degrees in $[S_{211}][3FPYRA]$, and 153 degrees in [S$_{211}$][BENZIM]. Based on the determined O…H distances and the obtuse O…H-N angles in the corresponding minimum point on the potential energy surface, the intermolecular H-bond formation can be declared as a result of chemisorption. Hydrogen bonding is an essential factor that makes $CO_2$ chemisorption an energetically favorable chemical reaction and shifts the chemisorption equilibrium rightward.

It is important to consider the redistribution of electron density over essential electrophilic and nucleophilic interaction centers once the $CO_2$ molecule grafts to the sulfonium-based ylide. The partial charges of the carboxylate oxygen atoms are -0.67e in [S$_{211}$][TRIAZ], -0.61e in [S$_{211}$][INDA], -0.67e in [S$_{211}$][3FPYRA], and -0.62e in [S$_{211}$][BENZIM]. The hydrogen charges in [H][AHA]s amount to +0.35e in [S$_{211}$][TRIAZ], +0.27e in [S$_{211}$][3FPYRA], +0.25e in [S$_{211}$][3FPYRA], and +0.38e in [S$_{211}$][BENZIM].

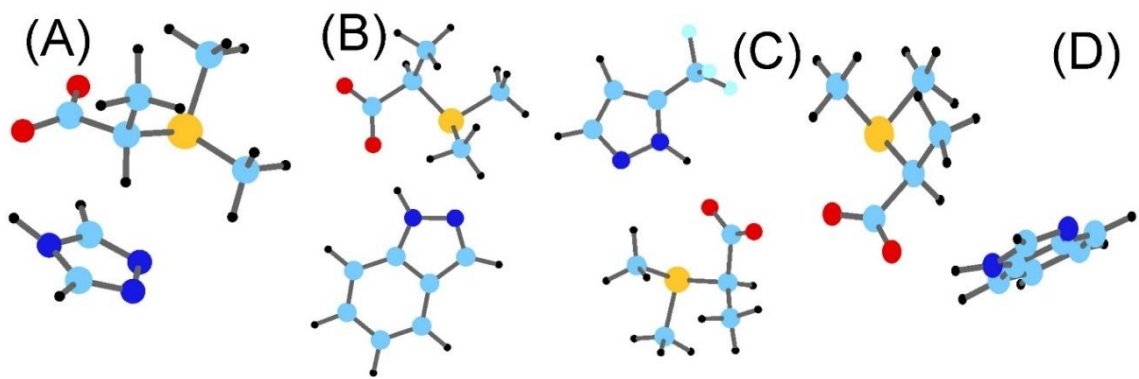

Figure 8. The structures of the sulfonium-based zwitterions and heterocyclic molecules obtained via the carboxylation reaction of the sulfonium-based ylide in various investigated RTILs: (A) [S$_{211}$][TRIAZ]; (B) [S$_{211}$][INDA], (C) [S$_{211}$][3FPYRA], and (D) [S$_{211}$][BENZ]. The carbon atoms are cyan, the hydrogen atoms are black, the nitrogen atoms are blue, the oxygen atoms are red, the fluorine atoms are light blue, and the sulfur atom is yellow.

The formation of the hydrogen bond leads to a substantial redistribution of the electron density over the heterocycle. For instance, the nitrogen atoms of all [H][AHA]s obtain more electron density as compared to the non-carboxylated configurations. The case of [S$_{211}$][3FPYRA] follows a general trend but stands somewhat aside due to the proximity of the trifluoromethyl moiety. The nitrogen partial charges are, therefore, routinely higher than in other species during the reaction. All partial charges on the chemisorption-relevant interaction centers of the sulfonium-based zwitterion and protonated heterocycles of the original RTILs are summarized in Table 1.

Table 1. The partial atomic charges of the $CO_2$ chemisorption-relevant atoms.

| Atom | Charge, e | | | |
|---|---|---|---|---|
| | [S$_{211}$][TRIAZ] | [S$_{211}$][INDA] | [S$_{211}$][3FPYRA] | [S$_{211}$][BENZIM] |
| S | +0.17 | +0.24 | +0.18 | +0.20 |
| O | -0.67 | -0.61 | -0.67 | -0.62 |
| H | +0.35 | +0.27 | +0.25 | +0.15 |
| α-C | +0.27 | +0.10 | +0.14 | +0.15 |
| N (H-bond) | -0.39 | -0.24 | +0.14 | -0.51 |
| N (no H-bond) | +0.25 | +0.21 | +0.40 | +0.27 |

The sulfur atom exhibits a fairly modest electrophilic behavior. The identity of [AHA] slightly influences the partial charge on sulfur. The most electrophilic sulfur atom was recorded in [S$_{211}$][INDA], +0.24e, whereas the least electrophilic sulfur was found in [S$_{211}$][TRIAZ], +0.17e. In turn, the α-carbon atom is most electrophilic in [S$_{211}$][TRIAZ], +0.27e, and least electrophilic in [S$_{211}$][INDA]. +0.10e. The two strongest attractions existing in the zwitterionic systems are (1) intramolecular oxygen-sulfur electrostatic coupling and (2) intermolecular O…H hydrogen bonding. Each of these paramount interactions involves the oxygen atom of the carboxyl group, i.e., the nucleophilic center of the zwitterion. Since both interactions are electrostatic, their strengths can be semiquantitatively assessed via the partial charges in Table 1 and relevant interatomic distances for each sorbent. The sulfur-oxygen distances within the carboxylated sulfonium-based zwitterion amount to 0.265 and 0.383 nm in [$S_{211}$][TRIAZ], 0.264 and 0.384 nm in [$S_{211}$][INDA], 0.260 and 0.381 nm in [$S_{211}$][3FPYRA], and 0.261 and 0.386 nm in [$S_{211}$][BENZIM]. We note that these intramolecular distances are smaller than the intermolecular hydrogen bonding discussed above. Since the partial charges are roughly similar on all involved interaction sites, the hydrogen bonds bring larger potential energy contributions.

**Carboxamidation of AHAs by carbon dioxide**

Irrespective of the sulfonium-based ylide carboxylation, the AHAs are capable of attaching $CO_2$ to their most nucleophilic interaction centers, i.e., nitrogen atoms. Figure 9 investigates the reaction coordinates for the chosen anions. Note that the counterion has been included in all simulations since the cation-anion electrostatic coupling is deemed important in the context of $CO_2$ capture. Within the above discussion, we showed that $CO_2$ must overcome the steric barrier before the carboxylation reaction takes place. This barrier is associated with the inherent location of the anion near the α-methylene group due to the electron deficiency of the sulfonium-based cation.

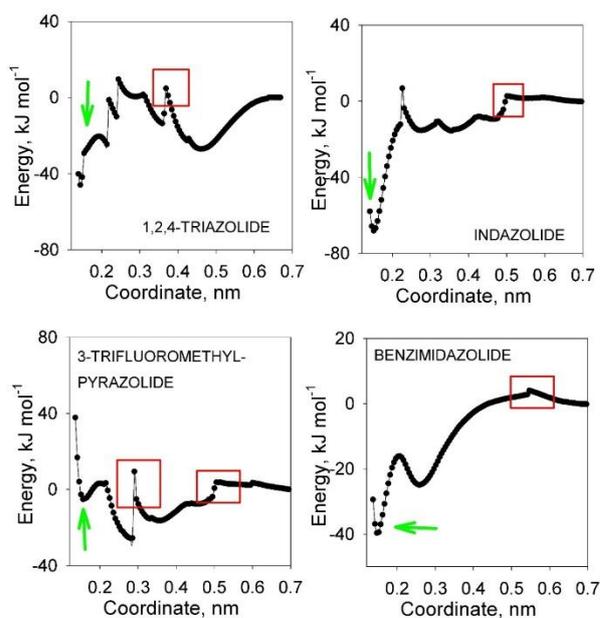

Figure 9. The evolutions of the reaction coordinates (nitrogen of anion-carbon of $CO_2$ distances) correspond to the carboxamidation formation reaction in the chosen ethyldimethylsulfonium RTILs. The red rectangles point to the steric barriers. The green arrows designate the states of the reaction products. The abrupt changes in energy correspond to the rotations of the reacting particles relative to one another.

The steric barriers were detected along the reaction coordinate. However, this kind of barrier is significantly smaller than those corresponding to the above-discussed carboxylation reaction. The detected heights of the highest steric barriers are 5 kJ/mol in $[S_{211}][TRIAZ]$, 3 kJ/mol in $[S_{211}][INDA]$, 10 kJ/mol in $[S_{211}][3FPYRA]$, and 5 kJ/mol in $[S_{211}][BENZIM]$. These barriers locate at drastically different anion-$CO_2$ distances, such as 0.37 nm in $[S_{211}][TRIAZ]$, 0.50 nm in $[S_{211}][INDA]$, 0.29 nm in $[S_{211}][3FPYRA]$, and 0.56 nm in $[S_{211}][BENZ]$. The barriers and their locations along the reaction coordinates represent complicated functions of reactive geometries and electronic densities in each system, therefore, it is scarcely possible to draw a simple trend. The barriers reflect the necessary reorientations of all particles within the reaction spot for the reaction coordinate to continue its movement. Numerous barriers in $[S_{211}][TRIAZ]$ correspond to the stepwise rotation of the [TRIAZ] AHA so that $CO_2$ meets the most nucleophilic nitrogen atom within the sorbent structure.

We have no experimental evidence that $CO_2$ performs carboxamidation necessarily at the most negatively charged nitrogen site. Furthermore, nucleophilicity adjusts stepwise as $CO_2$ approaches the AHA. The associated local structure changes substantially, and so are the nucleophilicities of the potential reaction sites. The simulation setup used herein is an obvious simplification and may need to be supplemented if more accurate knowledge becomes available, vide infra.

The chemisorption of $CO_2$ via carboxamidation of the anion is thermochemically allowed, though to various extents, depending on the chemical nature of the sample. The total energetic gains amount to -46 kJ/mol in [$S_{211}$][TRIAZ], -68 kJ/mol in [$S_{211}$][INDA], -6 kJ/mol in [$S_{211}$][3FPYRA], and -40 kJ/mol in [$S_{211}$][BENZIM]. The attachment of $CO_2$ to the most nucleophilic nitrogen atom of the AHA proceeds with an essentially negligible covalent barrier. However, the specific shape of the carboxamidation reaction coordinate differs in concordance with the chemical structure of each nucleophile (Figure 10). The small barriers identified along the anion's chemisorption coordinate qualitatively agree with the experimental data available to us. For instance, Brennecke and coworkers[33] reported chemisorption reactions by the same set of anions but paired them with the phosphonium-based cations. It was found that carboxamidation occurs readily and does not require any temperature elevation in contrast to the chemisorption by the mechanism of the phosphonium ylide.

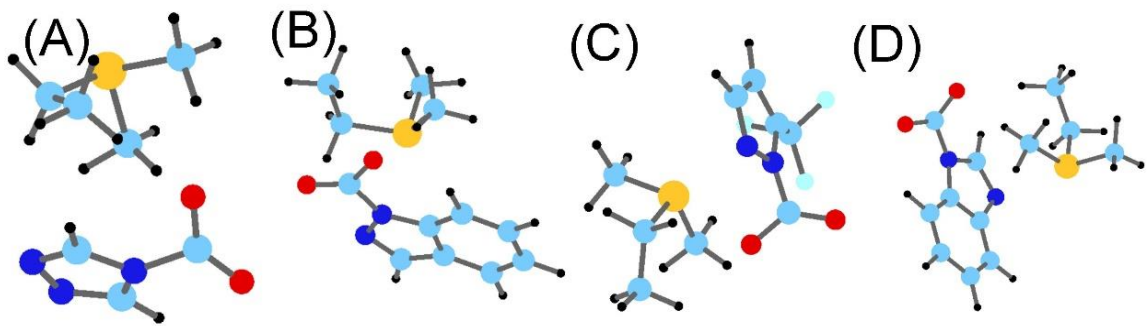

Figure 10. The structures of the $CO_2$ chemisorption products formed through the carboxamidation mechanism at the most electron-rich interaction centers of the simulated anions. The carbon atoms are cyan, the hydrogen atoms are black, the nitrogen atoms are blue, the oxygen atoms are red, the fluorine atoms are light blue, and the sulfur atoms are yellow.

While [$S_{211}$][TRIAZ], [$S_{211}$][INDA], and [$S_{211}$][BENZIM] exhibit the same type of behavior, [$S_{211}$][3FPYRA] stands out by a few distinctive features of its reaction profile. We found the connection between the poor thermochemistry of the $CO_2$ chemisorption reaction in [$S_{211}$][3FPYRA] and the electrostatic repulsion of the newly formed carboxyl moiety and the three fluorine atoms pertaining to the anion. For instance, the distance between the oxygen atom belonging to the carboxamido group and the closest fluorine atom amounts to as small as 0.281 nm. In turn, the partial charges on the fluorine atoms equal -0.15e, whereas the oxygen's partial charge is -0.60e. The electrostatic repulsion cannot be avoided with such a structural formula of the $CO_2$ chemisorption product. Therefore, the observed energetics is substantially less encouraging as compared to the other cases. While the trifluoromethyl moiety is important to stabilize a corresponding anionic species, it plays an adverse role during carboxamidation at the most nucleophilic nitrogen site. The energy profile for [$S_{211}$][3FPYRA] also contains a rather deep unexpected minimum at the reaction coordinate of 0.276 nm. The analysis of geometries and nucleophilicities at the point of interest unravels that the energy gain originates from the electrostatic coupling between the strongly electrophilic carbon atom of $CO_2$, +0.75e, and the nucleophilic fluorine atom, -0.19e, of the trifluoromethyl moiety which appeared at the distance of 0.293 nm before the carboxamidation. The covalent angle in $CO_2$ stays largely unperturbed, 176 degrees, at the discussion reaction coordinate. It indicates that the molecular configuration is far from the carboxamidation-associated saddle point.

To confirm the above-revealed peculiarity in [$S_{211}$][3FPYRA], we supplemented our simulation schedule to trial the carboxamidation of another nitrogen atom, i.e., a less nucleophilic reaction site. Figure 11 depicts an alternative path of $CO_2$ chemisorption by the [3FPYRA] anion. The carboxamidation of the N(1) site must suppress the intramolecular electrostatic repulsion and possibly adjust the force field in which $CO_2$ moves toward the target location. Indeed, both activation barrier and thermochemistry are more favorable for the N(1) and N(2) carboxamidations of [3FPYRA], 5 and -70 kJ/mol, respectively. The shape of the energy profile, particularly abrupt decreases of energy, reflects multiple rotations of the anion and $CO_2$ molecule. The on-the-fly rearrangements of the reactants bring about -20 kJ/mol to the system. The height of the covalent barrier is in line with other AHAs investigated. The visual inspection of the molecular conformations in the transition state and the product state unveiled no drastic steric hindrances unlike in the case of N(2). In turn, the proximity of the electron-rich moieties belonging to AHA to the core of the sulfonium-based cation stabilizes the products of $CO_2$ chemisorption. The mentioned distances amount to 0.355 nm from the carboxyl group to the sulfur atom and 0.312 nm from the trifluoromethyl group to the sulfur atom.

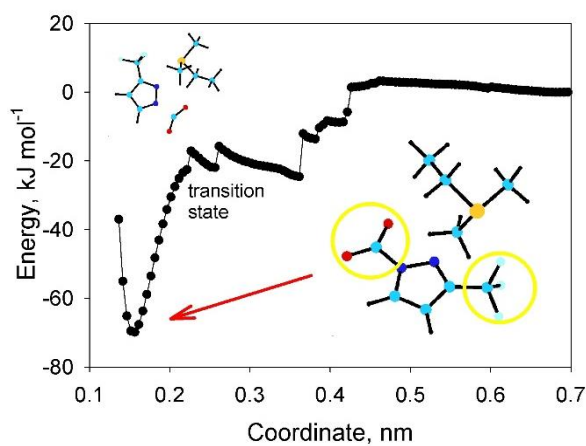

Figure 11. The evolution of the reaction coordinate (nitrogen of anion-carbon of $CO_2$ distance) corresponds to the carboxamidation reaction in [$S_{211}$][3FPYRA]. The chosen alternative carboxamidation site, N(1) is a weaker nucleophile than N(2) discussed above. The insets depict the product of $CO_2$ chemisorption and the transition state molecular geometry for the reaction. The abrupt changes in energy correspond to the rotations of the reacting particles relative to one another.

Our comprehensive analysis evidences that numerous factors are in action upon $CO_2$ chemisorption. First, the electrostatic attraction of the bond-forming sites, the nucleophilicity of nitrogen of AHA, and the electrophilicity of carbon of $CO_2$ matter. Second, the atomistic environment determined the forces that determined the flow of the chemical reaction. Third, the inherent geometrical rigidity imposes steric barriers. Since the geometries of the grafted groups and their orientations relative to the molecular core are predetermined by the hybridization states of the respective atoms, the prediction of the reaction profiles even in such simplified systems represents a fairly challenging endeavor. The obtained in-silico data help rationalize experimental observations to see the difference among various chemical structures of the AHAs.

**Sulfonium-based RTIL with non-AHA anion**

As we exemplified above, the ionic liquids composed of the ethyldimethylsulfonium cation and the cyclic anions exhibit promising performances as $CO_2$ scavengers. Even though the proton exchange stage is the limiting one, the emergence of carbamate and carboxylation make overall $CO_2$ chemisorption thermochemically favorable in all samples (Table 2). To verify whether the above-described behavior is unique to the aromatic structures, we computed the same anticipated reaction pathways for the [TFSI] anion paired with the [$S_{211}$] cation.

Figure 12 provides reaction energetic profiles, whereas Figure 13 visualizes a few representative molecular configurations along the chosen reaction coordinates. The proton exchange in [$S_{211}$][TFSI] is associated with a few steric points of extrema. These correspond to the rotations of the ions in the vicinity of one another to accommodate the alternating proton-nitrogen non-covalent distance. The recorded reaction profile does not exhibit a stationary point that would correspond to the protonated [TFSI] anion. The total potential energy of the system steeply grows when the reaction coordinate decreases below 0.190 nm. The reason for such an observation may lie in a three-dimensional geometry of the [TFSI]-anion and its symmetry. Indeed, the approach of the proton to the nitrogen atom of the anion assumes a simultaneous approach of various moieties of the cation (alkyl chains) and the anion (two sulfonyl groups) to one another. The nitrogen-sulfur covalent distances increase from 0.160 nm in the TFSI anion to 0.166 nm in the protonated (neutral) TFSI structure. The covalent sulfur-nitrogen-sulfur angle in TFSI increases from 121 degrees in the anionic form to 126 degrees in the neutral form. The α-carbon atom of the sulfonium-based ylide structure is located in tight proximity to the lost proton, 0.173 nm. The redistribution of the electron density results in the negative electrostatic charge on the migrating hydrogen atom (-0.26e) and the positive electrostatic charge on the α-carbon atom (+0.15e). Their attraction determines the unstable structures of the ethyldimethylsulfonium ylide and [H][TFSI].

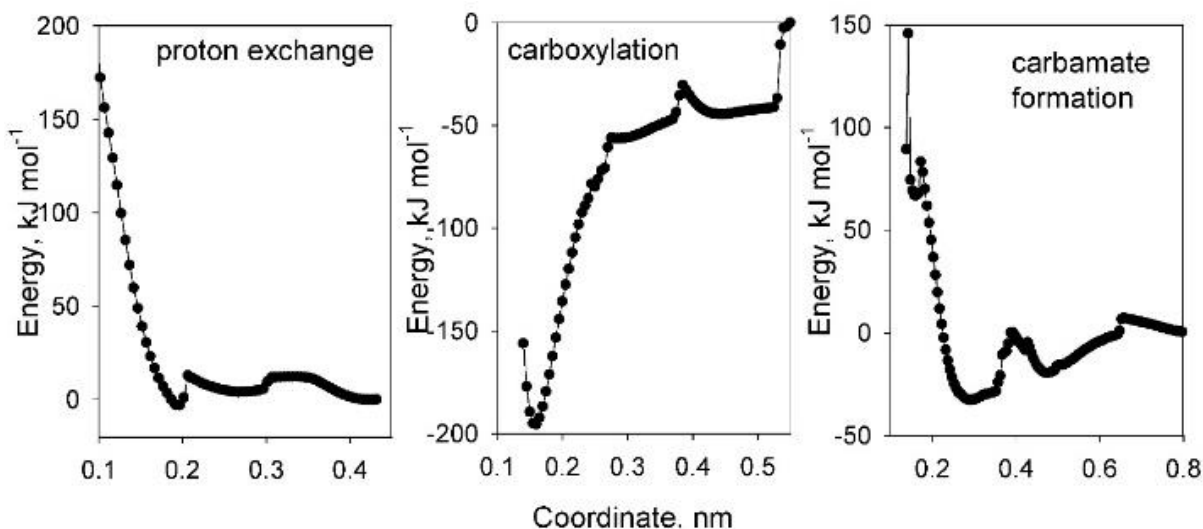

Figure 12. The $CO_2$ hypothetical chemisorption reactions by the $[S_{211}][TFSI]$ RTIL: proton exchange, carboxylation, and carbamate formation, see legends. The corresponding reaction coordinates are interatomic distances of hydrogen (cation)-nitrogen (anion), carbon (cation)-carbon ($CO_2$), and nitrogen (anion)-carbon ($CO_2$).

Several attempts were exercised to find the minimum stationary point for the $[S_{211}]$-ylide and [H][TFSI]. However, in vain. First, the reaction coordinate was scanned in reverse order, e.g., starting from [H][TFSI] toward the $[S_{211}]$ cation and [TFSI] anion. Second, the spatial step of the reaction coordinate scanning was decreased fivefold. Third, the geometry convergence criterion at every step along the minimum-energy pathway was tightened by an order of magnitude. In no case, the searched stationary point was located within the expected range of distances, 0.100-0.110 nm.

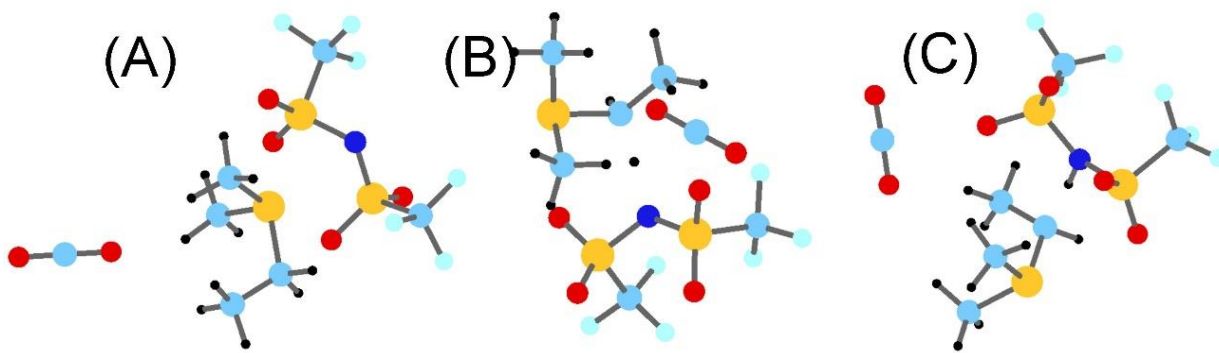

Figure 13. The proton transfer from the cation to the anion in ethyldimethylsulfonium bis(trifluoromethanesulfonyl)imide. (A) The spontaneous (minimum energy) molecular configuration of the ion pair and $CO_2$. (B) The transition state of the proton transfer reaction. (C) The unstable products are [$S_{211}$]-ylide and protonated bis(trifluoromethanesulfonyl)imide. The carbon atoms are cyan, the hydrogen atoms are black, the nitrogen atoms are blue, the oxygen atoms are red, the fluorine atoms are light blue, and the sulfur atom is yellow.

If the proton is forcibly transferred to the anion as characterized above, the carboxylation of the sulfonium-based ylide can be readily fulfilled. This reaction is strongly energetically favorable bringing nearly 200 kJ/mol of potential energy to the system. The carboxylated [$S_{211}$] cation strongly stabilizes by an electrostatic attraction between the protonated hydrogen atom and the newly formed carboxyl group.

Unlike in the cases of heterocyclic anions, the carboxamidation reaction is thermochemically forbidden with an energy effect of +67 kJ/mol. The barrier of this process is 85 kJ mol$^{-1}$. Brennecke and coworkers also probed $CO_2$ chemisorption by the [TFSI] anion coupled to the tetraalkylphosphonium cations. They found its unsatisfactory performance either at 333 or 353 K.[33] Based on the experiments, the basicity of the anion plays a paramount role in the context of deprotonation of the α-carbon atom of the cation alkyl chain. The actual deprotonation and subsequent formation of the ylide require larger basicity than the [TFSI] anion possesses. According to ATR-FTIR and $^{31}$P NMR, the deprotonation of the phosphonium-based cations did not take place, whereas we hereby confirm and corroborate their results using the sulfonium-based cation.[33] Indeed, strong enough hydrogen bonds cannot form between the relatively weak bases and the relatively non-polar proton of the α-carbon atom.

The failure of the [TFSI] anion to compete with the AHAs can be readily correlated to the energies of the valence electrons in the systems. The HOMO level in [$S_{211}$][TFSI] has an energy of -9.9 eV which is significantly lower compared to all heterocyclic structures, e.g., -6.7 eV in [$S_{211}$][TFSI], -7.0 eV in [$S_{211}$][BENZIM], -7.8 eV in [$S_{211}$][TRIAZ], and -8.0 eV in [$S_{211}$][3FPYRA]. The low energies of the valence electrons of the reactants imply the high stabilities of the corresponding electron-nuclei systems. They make most chemical reactions thermochemically unfavorable. The [TFSI] anion represents a rather stable chemical entity whose protonation by α-methylene hydrogen and carboxamidation by $CO_2$ are not within favorable processes.

**Energy profiles in light of the structure of AHAs**

The energetics of the chemical reaction taking place between many-atom reagents depends on many electronic and structural peculiarities of the initial, transition, and final state of the molecules. Because lots of factors are simultaneously and consequently involved, neither the barrier nor thermochemistry can be reliably hypothesized from scratch. Noteworthy, the deprotonation activation barriers only pseudorandomly correlate with the thermochemistry. For instance, in [$S_{211}$][TRIAZ], [$S_{211}$][3FPYRA], and [$S_{211}$][BENZIM], the lower barriers correspond to the more favorable thermochemistry. However, in [$S_{211}$][INDA], the most favorable thermochemistry coexists with the most unfavorable barrier. The carboxylation of [$S_{211}$] proceeds similarly according to the thermochemistry but differs severely according to the activation energies. Recall the highest barrier in [$S_{211}$][3FPYRA] and the lowest barrier in [$S_{211}$][INDA], whereas the energetic effects of both reactions are equal. The carboxamidation of the anion occurs expectedly straightforwardly in all [$S_{211}$][AHA]s, whereas the thermochemistry of each $CO_2$ capture differs substantially. The [INDA] anion exhibits the best energetics of carboxamidation at the most nucleophilic site of the anion. This feature positions [S$_{211}$][INDA] as the most thermochemically favored CO$_2$ scavenger out of the investigated dataset.

Table 2. The summary of thermochemical quantities characterizing the CO$_2$ chemisorption reaction by the investigated [S$_{211}$]-based RTILs. ΔU designates the energetic effect of the reaction, whereas ΔA designates the height of the covalent activation barrier. Note that the activation barriers are measured relative to the closest minimum along the respective reaction coordinate. In turn, the reaction profiles are given relative to the stationary point configurations of the reactants.

| Sorbent | Energy, kJ/mol | | | | | |
|---|---|---|---|---|---|---|
| | ΔU$^{deprot}$ | ΔA$^{deprot}$ | ΔU$^{cation}$ | ΔA$^{cation}$ | ΔU$^{anion}$ | ΔA$^{anion}$ |
| [S$_{211}$][TRIAZ] | +50 | +60 | -109 | +26 | -46 | +5 |
| [S$_{211}$][INDA] | +13 | +61 | -100 | +20 | -68 | +3 |
| [S$_{211}$][3FPYRA] | +27 | +47 | -100 | +52 | -70 | +3 |
| [S$_{211}$][BENZIM] | +21 | +38 | -105 | +23 | -40 | +5 |
| [S$_{211}$][TFSI] | ∞ | +14 | -196 | 0 | +67 | +85 |

[S$_{211}$][TFSI] does not appear competitive to the [S$_{211}$][AHA]s in the context of CO$_2$ capture. The [TFSI] anion does not foster the ethyldimethylsulfonium ylide formation and does not undergo carboxamidation at its nitrogen site. Since the carboxylation of [S$_{211}$] depends on the preceding deprotonation, this stage does not happen in [S$_{211}$][TFSI]. In a hypothetical case where the carboxylation was realistic, the associated energy gain would amount to an outstanding value of -196 kJ/mol.

The comparisons of the HOMO energy levels of the reactants for the same-type reactions permit one to assess their likelihood. For instance, the highest HOMO level is observed in [S$_{211}$][INDA] and corresponds to the most straightforward CO$_2$ capture by the indazolide anion. In turn, the lowest HOMO observed in [S$_{211}$][3FPYRA] corresponds to the least favorable carboxamidation of [3FPYRA].

**Conclusions and final remarks**

The feasibility of $CO_2$ chemisorption by a hypothetical class of RTILs composed of the $[S_{211}]$ cation and a set of AHAs was characterized via reaction profiles, evolving atomic nucleophilicities, geometrical alterations, and relevant molecular properties. The electron-rich sites of the AHAs exhibit high and differing proton affinities to perform the deprotonation of the α-carbon atom of the sulfonium-based cation. The ethyldimethylsulfonium ylide emerges and exists in equilibrium with the $[S_{211}]$-based RTILs. The subsequent carboxylation of the sulfonium α-carbon atom by $CO_2$ results in the emergence of the carboxylate moiety. The negatively charged carboxylate moiety electrostatically attracts the central atom of the precursor sulfonium-based cation getting in this way stabilized. We identified in this work, for the first time, that the deprotonation of the sulfonium-based cation by the AHAs is associated with moderate energy losses, 13-50 kJ/mol, and commensurate activation barriers. Such a process is, therefore, expected to occur at somewhat elevated temperatures. In turn, the $CO_2$ chemisorption stage (carboxylation) excessively compensates for deprotonation-related energy losses. The $[S_{211}]$-ylide is quite thermodynamically unstable and, therefore, represents a chemically reactive compound. On the aggregate, the carbon dioxide capture by the sulfonium-based cations should be considered a relatively straightforward process.

For comparison, the investigation of $[S_{211}][TFSI]$ revealed that an acyclic anion [TFSI] does not favor the carboxylation of the sulfonium-based cation. Such a reaction gets blocked at the ylide formation stage. [TFSI] is also not a competitive $CO_2$ scavenger itself. Therefore, the principal difference between cyclic aromatic and acyclic non-aromatic anions was recorded in the context of $CO_2$ sorption and linked to the nucleophilicities of the involved in-ring nitrogen interaction sites.

The nitrogen-containing AHAs participate in $CO_2$ chemisorption following the carboxamidation pathway. The reaction takes place at the most nucleophilic interaction center unless structural peculiarities alter such a behavior, see the case of [$S_{211}$][3FPYRA]. This reaction route contains one or more steric barriers (up to 10 kJ/mol) and insignificant covalent activation barriers (< 20 kJ/mol). The steric barriers relate to the approach of the solvated $CO_2$ molecule to the target reaction site. We observed that the $CO_2$-anion separation in the physisorbed molecular configurations of the reactants is over 0.7 nm.

In the physisorbed configuration of $CO_2$ at [$S_{211}$][AHA]s, the most electron-rich center of the anion coordinates the most electron-deficient center of the cation, whereas $CO_2$ locates relatively far from both of them. Since no stationary point of the reactants includes $CO_2$ tightly coordinated by any ion, we hypothesize that the presence of the gas molecules is unessential for the proton transfer reaction. During proton migration, the ion-molecular geometries of the systems somewhat adjust to reflect the conversion of ions into molecules. The latter naturally leads to lowering intermolecular electrostatic forces.

Chemisorption is made thermochemically possible thanks to the two geometrical features of the products. The hydrogen bonds emerging between the newly acquired carboxyl moiety of [$S_{211}$] and the protonating hydrogen atom of the heterocycle play an important role in the stabilization of captured $CO_2$. The additional stabilization comes from the electrostatic attraction between the center of the cation and its carboxyl moiety. Since the electron density is strongly delocalized over the ions, the α-carbon atom appears somewhat electron deficient. The electrostatic attraction between them and the oxygen atoms of the carboxylate group occurring at small distances represents an important factor that determines the ultimate reaction energetics.

To recapitulate, the sulfonium-based RTILs were numerically detected to capture $CO_2$ engaging both the cation and the anion. Two sorbate molecules can be fixed by one mole of a sorbent, theoretically. Both processes involve the formation of chemical bonds and both are thermochemically favored. The heights of the detected activation barriers are mediocre and can be fine-tuned by the AHAs employed. The proved principle urges experimental validation.

**Conflict of interest**

The author hereby declares no financial interests that might bias the interpretations of the obtained results.

**Author for correspondence**

Inquiries regarding the scientific content of this paper shall be directed through electric mail to Prof. Dr. Vitaly V. Chaban (vvchaban@gmail.com).

**References**

1. Chaban, V. V.; Andreeva, N. A. Combating Global Warming: Moderating Carbon Dioxide Concentration in the Earth's Atmosphere through Robust Design of Novel Scavengers. LAP LAMBERT Academic Publishing, **2018**, 164 pages. isbn: 9786137327098.
2. Wang, C.; Lv, Z.; Yang, W.; Feng, X.; Wang, B. A rational design of functional porous frameworks for electrocatalytic CO2 reduction reaction. Chemical Society Reviews, **2023**, 10.1039/d2cs00843b.

3. Song, K. S.; Fritz, P. W.; Coskun, A. Porous organic polymers for CO2 capture, separation and conversion. Chemical Society Reviews, **2022**, 51 (23), 9831-9852, 10.1039/d2cs00727d.

4. Fu, D.; Davis, M. E. Carbon dioxide capture with zeotype materials. Chemical Society Reviews, **2022**, 51 (22), 9340-9370, 10.1039/d2cs00508e.

5. Halder, A. K.; Ambure, P.; Perez-Castillo, Y.; Cordeiro, M. N. D. S. Turning deep-eutectic solvents into value-added products for CO2 capture: A desirability-based virtual screening study. Journal of CO2 Utilization, **2022**, 58, 101926, 10.1016/j.jcou.2022.101926.

6. Carrera, G.; Inês, J.; Bernardes, C. E. S.; Klimenko, K.; Shimizu, K.; Canongia Lopes, J. N. The Solubility of Gases in Ionic Liquids: A Chemoinformatic Predictive and Interpretable Approach. ChemPhysChem, **2021**, 22 (21), 2190-2200, 10.1002/cphc.202100632.

7. Ferreira, T. J.; de Moura, B. A.; Esteves, L. M.; Reis, P. M.; Esperança, J. M. S. S.; Esteves, I. A. A. C. Biocompatible ammonium-based ionic liquids/ZIF-8 composites for CO2/CH4 and CO2/N2 separations. Sustainable Materials and Technologies, **2023**, 35, e00558, 10.1016/j.susmat.2022.e00558.

8. Hussain Solangi, N.; Hussin, F.; Anjum, A.; Sabzoi, N.; Ali Mazari, S.; Mubarak, N. M.; Kheireddine Aroua, M.; Siddiqui, M. T. H.; Saeed Qureshi, S. A review of encapsulated ionic liquids for CO2 capture. Journal of Molecular Liquids, **2023**, 374, 121266, 10.1016/j.molliq.2023.121266.

9. Hernández, E.; Santiago, R.; Moya, C.; Navarro, P.; Palomar, J. Understanding the CO2 valorization to propylene carbonate catalyzed by 1-butyl-3-methylimidazolium amino acid ionic liquids. Journal of Molecular Liquids, **2021**, 324, 114782, 10.1016/j.molliq.2020.114782.

10. Andreeva, N. A.; Chaban, V. V. Amino-functionalized ionic liquids as carbon dioxide scavengers. Ab initio thermodynamics for chemisorption. Journal of Chemical Thermodynamics, **2016**, 103, 1-6, 10.1016/j.jct.2016.07.045.

11. Andreeva, N. A.; Chaban, V. V. Electronic and thermodynamic properties of the amino- and carboxamido-functionalized C-60-based fullerenes: Towards non-volatile carbon dioxide scavengers. Journal of Chemical Thermodynamics, **2018**, 116, 1-6, 10.1016/j.jct.2017.08.019.

12. Chaban, V. V.; Andreeva, N. A. Extensively amino-functionalized graphene captures carbon dioxide. Physical Chemistry Chemical Physics, **2022**, 24 (42), 25801-25815, 10.1039/d2cp03235j.

13. Kaur, G.; Kumar, H.; Singla, M. Diverse applications of ionic liquids: A comprehensive review. Journal of Molecular Liquids, **2022**, 351, 118556, 10.1016/j.molliq.2022.118556.

14. Chaban, V. V.; Prezhdo, O. V. Ionic and molecular liquids: Working together for robust engineering. Journal of Physical Chemistry Letters, **2013**, 4 (9), 1423-1431, 10.1021/jz400113y.

15. Zhuang, W.; Hachem, K.; Bokov, D.; Javed Ansari, M.; Taghvaie Nakhjiri, A. Ionic liquids in pharmaceutical industry: A systematic review on applications and future perspectives. Journal of Molecular Liquids, **2022**, 349, 118145, 10.1016/j.molliq.2021.118145.

16. Esperança, J. M. S. S.; Canongia Lopes, J. N.; Tariq, M.; Santos, L. M. N. B. F.; Magee, J. W.; Rebelo, L. P. N. Volatility of aprotic ionic liquids - A review. Journal of Chemical and Engineering Data, **2010**, 55 (1), 3-12, 10.1021/je900458w.

17. Voroshylova, I. V.; Ers, H.; Koverga, V.; Docampo-Álvarez, B.; Pikma, P.; Ivaništšev, V. B.; Cordeiro, M. N. D. S. Ionic liquid–metal interface: The origins of capacitance peaks. Electrochimica Acta, **2021**, 379, 138148, 10.1016/j.electacta.2021.138148.
18. Haghani, H.; Behrouz, M.; Chaban, V. V. Triethylsulfonium-based ionic liquids enforce lithium salt electrolytes. Physical Chemistry Chemical Physics, **2022**, 24 (16), 9418-9431, 10.1039/d2cp00275b.
19. Khachatrian, A. A.; Shamsutdinova, Z. I.; Rakipov, I. T.; Varfolomeev, M. A. The ability of ionic liquids to form hydrogen bonds with organic solutes evaluated by different experimental techniques. Part I. Alkyl substituted imidazolium and sulfonium based ionic liquids. Journal of Molecular Liquids, **2018**, 265, 238-242, 10.1016/j.molliq.2018.05.136.
20. Deyab, M. A. Sulfonium-based ionic liquid as an anticorrosive agent for thermal desalination units. Journal of Molecular Liquids, **2019**, 296, 111742, 10.1016/j.molliq.2019.111742.
21. Chaban, V. V.; Andreeva, N. A.; Voroshylova, I. V. Ammonium-, phosphonium- and sulfonium-based 2-cyanopyrrolidine ionic liquids for carbon dioxide fixation. Physical Chemistry Chemical Physics, **2022**, 24 (16), 9659-9672, 10.1039/d2cp00177b.
22. Medeiros, L. H. G. D.; Alves, A. A. A.; Feitosa, F. X.; Sant'Ana, H. B. D. Cation effect on bis(trifluoromethylsulfonyl)imide-based ionic liquids with triethylsulfonium, 1,2-dimethyl-3-propylimidazolium, 1-methyl-1-propylpyrrolidinium, and 1-butyl-2,3-dimethylimidazolium density at high pressure. Journal of Molecular Liquids, **2022**, 354, 118851, 10.1016/j.molliq.2022.118851.
23. Chaban, V. V. Carbon Dioxide Chemisorption by Ammonium and Phosphonium Ionic Liquids: Quantum Chemistry Calculations. Journal of Physical Chemistry B, **2022**, 126 (29), 5497-5506, 10.1021/acs.jpcb.2c02968.
24. Peverati, R.; Truhlar, D. G. Performance of the M11 and M11-L density functionals for calculations of electronic excitation energies by adiabatic time-dependent density functional theory. Physical Chemistry Chemical Physics, **2012**, 14 (32), 11363-11370, 10.1039/C2CP41295K.
25. Risthaus, T.; Grimme, S. Benchmarking of London Dispersion-Accounting Density Functional Theory Methods on Very Large Molecular Complexes. Journal of Chemical Theory and Computation, **2013**, 9 (3), 1580-1591, 10.1021/ct301081n.
26. Singh, U. C.; Kollman, P. A. An approach to computing electrostatic charges for molecules. Journal of Computational Chemistry, **1984**, 5 (2), 129-145, 10.1002/jcc.540050204.
27. Schmidt, M. W.; Baldridge, K. K.; Boatz, J. A.; Elbert, S. T.; Gordon, M. S.; Jensen, J. H.; Koseki, S.; Matsunaga, N.; Nguyen, K. A.; Su, S.; Windus, T. L.; Dupuis, M.; Montgomery, J. A. General atomic and molecular electronic structure system. Journal of Computational Chemistry, **1993**, 14 (11), 1347-1363, 10.1002/jcc.540141112.
28. Hanwell, M. D.; Curtis, D. E.; Lonie, D. C.; Vandermeersch, T.; Zurek, E.; Hutchison, G. R. Avogadro: an advanced semantic chemical editor, visualization, and analysis platform. Journal of Cheminformatics, **2012**, 4 (1), 17, 10.1186/1758-2946-4-17.
29. Allouche, A. R. Gabedit-A Graphical User Interface for Computational Chemistry Softwares. Journal of Computational Chemistry, **2011**, 32 (1), 174-182, 10.1002/jcc.21600.
30. Virtanen, P.; Gommers, R.; Oliphant, T. E.; Haberland, M.; Reddy, T.; Cournapeau, D.; Burovski, E.; Peterson, P.; Weckesser, W.; Bright, J.; van der Walt, S. J.; Brett, M.; Wilson, J.; Millman, K. J.; Mayorov, N.; Nelson, A. R. J.; Jones, E.; Kern, R.; Larson, E.; Carey, C. J.; Polat, I.; Feng, Y.; Moore, E. W.; VanderPlas, J.; Laxalde, D.; Perktold, J.;

Cimrman, R.; Henriksen, I.; Quintero, E. A.; Harris, C. R.; Archibald, A. M.; Ribeiro, A. H.; Pedregosa, F.; van Mulbregt, P.; SciPy, C. SciPy 1.0: fundamental algorithms for scientific computing in Python. Nat Methods, **2020**, 17 (3), 261-272, 10.1038/s41592-019-0686-2.

31. Bahn, S. R.; Jacobsen, K. W. An object-oriented scripting interface to a legacy electronic structure code. Computing in Science & Engineering, **2002**, 4 (3), 56-66, 10.1109/5992.998641.

32. Chaban, V. V.; Prezhdo, O. V. Energy Storage in Cubane Derivatives and Their Real-Time Decomposition: Computational Molecular Dynamics and Thermodynamics. ACS Energy Letters, **2016**, 1 (1), 189-194, 10.1021/acsenergylett.6b00075.

33. Gohndrone, T. R.; Song, T.; DeSilva, M. A.; Brennecke, J. F. Quantification of Ylide Formation in Phosphonium-Based Ionic Liquids Reacted with CO2. Journal of Physical Chemistry B, **2021**, 125 (24), 6649-6657, 10.1021/acs.jpcb.1c03546.

34. Chaban, V. V.; Andreeva, N. A. Structure, thermodynamic and electronic properties of carbon-nitrogen cubanes and protonated polynitrogen cations. Journal of Molecular Structure, **2017**, 1149, 828-834, 10.1016/j.molstruc.2017.08.063.